\documentstyle[12pt]{article}
\newtheorem{Thm}{Theorem}[section]

\newtheorem{Prop}[Thm]{Proposition}
\newtheorem{Ass}[Thm]{Assumption}
\newenvironment{proof}{\noindent {\bf proof:}\newline}{\QED}
\renewcommand{\theequation}{\mbox{\arabic{section}.\arabic{equation}}}
\newcounter{saveeqn}
\newcommand{\alpheqn}{\setcounter{saveeqn}{\value{equation}}%
  \stepcounter{saveeqn}\setcounter{equation}{0}%
  \renewcommand{\theequation}{%
  \mbox{\arabic{section}.\arabic{saveeqn}-\alph{equation}}}}
\newcommand{\reseteqn}{\setcounter{equation}{\value{saveeqn}}%
\renewcommand{\theequation}{\mbox{\arabic{section}.\arabic{equation}}}}
\newcommand{\newsection}[1]{\section{#1}\setcounter{equation}{0}}
\newcommand{\eq}[1]{(\ref{#1})}

\newcommand{\be}{\begin{equation}}
\newcommand{\ee}{\end{equation}}
\newcommand{\bea}{\begin{eqnarray}}
\newcommand{\eea}{\end{eqnarray}}
\newcommand{\beann}{\begin{eqnarray*}}
\newcommand{\eeann}{\end{eqnarray*}}

\newcommand{\Hs}{{\cal H}}

\newcommand{\up}{\uparrow}
\newcommand{\down}{\downarrow}
\newcommand{\Gammalower}{\Gamma_{\rm low}}
\newcommand{\Gammaupper}{\Gamma_{\rm upp}}
\newcommand{\ket}[1]{\left\vert #1\right\rangle}
\newcommand{\bra}[1]{\left\langle #1\right\vert}

\def\idty{{\leavevmode{\rm 1\ifmmode\mkern -5.4mu\else
    \kern -.3em\fi I}}}
\def\Ibb #1{ {\rm I\ifmmode\mkern -3.6mu\else\kern -.2em\fi#1}}
\def\Ird{{\hbox{\kern2pt\vbox{\hrule height0pt depth.4pt width5.7pt
    \hbox{\kern-1pt\sevensy\char"36\kern2pt\char"36} \vskip-.2pt
    \hrule height.4pt depth0pt width6pt}}}}
\def\Irs{{\hbox{\kern2pt\vbox{\hrule height0pt depth.34pt width5pt
    \hbox{\kern-1pt\fivesy\char"36\kern1.6pt\char"36} \vskip -.1pt
    \hrule height .34 pt depth 0pt width 5.1 pt}}}}
\def\Ir{\mbox{\bf Z}}
\def\ibb #1{\leavevmode\hbox{\kern.3em\vrule
     height 1.5ex depth -.1ex width .2pt\kern-.3em\rm#1}}
 \def\Cx {{\ibb C}} 

\def\QED{\\ {\hspace*{\fill}{\vrule height 1.8ex width 1.8ex }\quad}
    \vskip 0pt plus20pt}

\def\A1n{A_1\otimes\cdots\otimes A_n}

\def\dom{\mathop{\rm dom}\nolimits}

\def\ker{{\rm ker}\,}

\def\ran{{\rm ran}\,}

\def\Tr{\mathop{\rm Tr}\nolimits}

\def\phi{\varphi}            
\def\epsilon{\varepsilon}    

\def\A{{\cal A}}   
 \def\H{{\cal H}}

\catcode`@=11
\def\@biblabel#1{#1.}
\def\ifundefined#1{\expandafter\ifx\csname
                        \expandafter\eat\string#1\endcsname\relax}
\def\atdef#1{\expandafter\def\csname #1\endcsname}
\def\atedef#1{\expandafter\edef\csname #1\endcsname}
\def\atname#1{\csname #1\endcsname}
\def\ifempty#1{\ifx\@mp#1\@mp}
\def\ifatundef#1#2#3{\expandafter\ifx\csname#1\endcsname\relax
                                  #2\else#3\fi}
\def\eat#1{}
\newdimen\refskip  \refskip=0pt 
\def\@utfirst #1,#2\@ver
   {\author#1,\ifx#2\@ut\afteraut\else\@utsecond#2\@ver\fi}
\def\@utsecond #1,#2\@ver
   {\ifx#2\@ut\andone\author#1,\afterauts\else
      ,\author#1,\@utmore#2\@ver\fi}
\def\@utmore #1,#2\@ver
   {\ifx#2\@ut\and\author#1,\afterauts\else
      ,\author#1,\@utmore#2\@ver\fi}
\def\authors#1{\@utfirst#1,\@ut\@ver}
\let\more\relax  
\catcode`@=12
\def\Bref#1 "#2"#3{\authors{#1}:\ {\it #2}, #3\more}
\def\Gref#1 "#2"#3{\authors{#1}\ifempty{#2}\else:``#2''\fi, #3\more}
\def\Grefnt#1 "#2"#3{\authors{#1}, #3\more} 
\def\Jref#1 "#2"#3{\relax \authors{#1}:``#2'', \Jn{#3}}
\def\Jrefnt#1 "#2"#3{\relax \authors{#1}, \Jn{#3}} 
\def\inPr#1 "#2"#3\relax{in: \authors{\eds#1}:{\it #2}, #3}
\newcommand{\Jn}[4]{{\it#1}\ {\bf#2}, #4 (#3)}
\def\author#1. #2,{#1.~#2}
\def\sameauthor#1{\leavevmode$\underline{\hbox to 25pt{}}$}
\def\and{, and}   \def\andone{ and}
\def\noinitial#1{\ignorespaces}
\let\afteraut\relax
\let\afterauts\relax
\def\etal{\def\afteraut{, et.al.}\let\afterauts\afteraut
           \let\and,}
\def\eds{\def\afteraut{(ed.)}\def\afterauts{(eds.)}}
\newcommand{\condmat}[1]{archived as {\tt cond-mat/#1}}
\begin{document}
\thispagestyle{empty}
{Archived as {\tt cond-mat/9501098 }\hspace{\fill} Preprint BN-16Jan95}
\vspace{30pt}
\begin{center}
{\LARGE\bf Bounds on the mass gap of the} \\[10pt]
{\LARGE\bf ferromagnetic XXZ chain} \\[30pt]
Bruno Nachtergaele\\
Department of Physics\\
Princeton University\\
Princeton, NJ 08544-0708, USA\\
E-mail: {\tt bxn@math.princeton.edu}\\[20pt]
(16 January 1995)\\[40pt]

{\bf Abstract}\\[20pt]
\end{center}
We prove rigorous lower and upper bounds for the mass gap of the
ferromagnetic spin 1/2 XXZ chain. The two bounds coincide
asymptotically in the Ising limit $\Delta\to\infty$. Near the
isotropic point, $\Delta=1$, the estimates are good enough to
determine the critical behaviour of the mass gap unambiguously.
The derivation does not rely on exact solutions.

\vspace{20pt}

\vfill

\hrule width2truein
\smallskip
{\baselineskip=12pt
\noindent
Copyright \copyright\ 1995 by the author. Faithful reproduction of
this article by any means is permitted for non-commercial purposes.\par
}
\newpage

\newsection{Introduction}\label{sec:intro}

The main purpose of this paper is to establish rigorous upper and lower
bounds on the spectral gap of the one-dimensional ferromagnetic spin
1/2 XXZ Heisenberg chain.  See below in Theorem \ref{thm:main} for the
explicit formulae.  The upper bound is generally assumed to be the
exact result. The lower bound has essentially the same behaviour and
is reasonably close to the upper bound.  It is good enough to
determine unambiguously the critical exponent $\alpha$ of the model.
We hope to make clear that our method relies only on certain
properties of the ferromagnetic XXZ chain and not on the exact
solution.  We expect that the method will be useful in the study of
other models that share the same general properties without being
exactly soluble. In fact, similar ideas were already applied to a
class of quantum spin chains with discrete symmetry breaking in
\cite{GVBS}. We refer the reader to Section \ref{sec:gap} for a
discussion of the method and further references.

The ferromagnetic regime of the XXZ chain has not been studied as
extensively as the antiferromagnetic and the critical regimes. One
could think that this is because it is less interesting or less
challenging. We think this is only partly correct. In fact, the
literature makes it very clear that the model is not so well
understood as one might deduce from the fact that it is ``Bethe Ansatz
soluble''. The more careful practitioners of the art do not neglect to
point out that the validity of the Bethe Ansatz solution relies on the
so-called string hypothesis, which remains unproven till now. In fact,
it is known that the string hypothesis cannot universally hold, i.e.,
it is violated for some finite chains \cite{Sch}.

Proofs of the completeness of the Bethe Ansatz based on combinatorial
arguments (counting the number of solutions of the Bethe equations)
always assume the string hypothesis \cite{Kir}. The proof of Yang and
Yang \cite{YY} of the validity of the Bethe Ansatz for the ground
state of the XXZ chain only covers the range $-\infty< \Delta <-1$,
which is the complement of the regime studied in the present work. For
the case $\Delta =1$ a form of completeness in the thermodynamic limit
was shown in \cite{BT}.

In view of the general uniqueness theorem \cite{Ara} for the
Gibbs state of one-dimensional quantum spin models with
translation invariant finite range interactions, there should
be no surprises in the finite temperature behaviour of the XXZ
chain. Yet, there are still unresolved questions about the low
temperature behaviour of the specific heat of the model in the
ferromagnetic region (See \cite{Joh} and \cite[Chapter 6]{Mat}
for a discussion.)

These unresolved questions clearly demonstrate the need for rigorous
arguments. The arguments presented in this paper do not address all of
them, but they provide unambiguous information on the behaviour of
the mass gap.

The XXZ-Hamiltonian for a finite chain of $L$ sites, including
the special boundary conditions that we consider, is
\begin{equation}
H^{XXZ}_L=A(\Delta)(S_L^3-S_1^3)
-\sum_{x=1}^{L-1} \frac{1}{\Delta}(S^1_xS^1_{x+1}+S^2_xS^2_{x+1})
+S^3_xS^3_{x+1}
\label{ham+bc}
\end{equation}
where $S_x^\alpha,\alpha=1,2,3$, are the usual $2\times 2$ spin
matrices (with eigenvalues $\pm 1/2$) acting on the site $x$, and
$A(\Delta)=\pm {1\over 2}\sqrt{1-1/\Delta^2}$. The two Hamiltonians
corresponding to the positive and negative choice of $A(\Delta)$ are
obviously unitarily equivalent by left-right symmetry. Unless
explicitly mentioned we will always refer to the Hamiltonian with the
positive choice for $A(\Delta)$. The boundary conditions and
normalization of \eq{ham+bc} are natural for the following reasons.
First of all they make the ground state degeneracy equal to $L+1$ {\it
for all $\Delta\geq 1$}. This property can be explained in terms of a
quantum group symmetry that the Hamiltonian, with these particular
boundary terms included, possesses \cite{PS}. The normalization is
such that one can consider the limit $\Delta\to\infty$ without
difficulty. In this limit the model becomes the ferromagnetic Ising
chain with a boundary term that allows for ground states with a kink,
i.e., for any site $x$ in the finite chain the configuration with all
spins to the left of $x$ {\it up \/} ($\up$) and  all spins to the
right of $x$ {\it down\/} ($\down$), is a ground state. Obviously
there are $L-1$ of such kink states. Together with the two translation
invariant configurations this yields $L+1$ ground states.
As we will see in Section \ref{sec:infinite} the
boundary terms also make the computation of the GNS Hamiltonians
of the infinite chain immediate.

In the isotropic limit ($\Delta=1$) the $L+1$-fold degeneracy is the
dimension of the spin $L/2$ representation of SU(2). Note that the
boundary terms vanish for $\Delta=1$. In the thermodynamic limit
($L\to\infty$) all translation invariant ground states are
states of perfectly aligned spins. No non translation invariant
ground states are known.

If $\Delta >1$, there are four different classes of known ground states
of the model on the infinite chain, which could be called {\it up,
down, kink\/}, and {\it antikink\/}.  They consist, respectively, of
the state with all spins $\up$, the state with all spins $\down$, an
infinite number of states in which the spins are $\up$ at $-\infty$
and $\down$ at $+\infty$, and an infinite number of states in which
the spins are $\down$ at $-\infty$ and $\up$ at $+\infty$. The
infinite degeneracy of the ground state in the latter two sectors
corresponds to the possible choices for the location of the kink or
antikink, i.e., the location on the chain where the spins turn over
form up to down in the first and down to up in the second case.  The
kinks are strictly speaking located at a single bond only in the Ising
limit ($\Delta\to\infty$).  For $1<\Delta<\infty$ the ground states
are not described by a single configuration because of the quantum
fluctuations, but the kinks, respectively antikinks, are
quasilocalized for all $1<\Delta<\infty$.  We refer to Section
\ref{sec:ground} for more explicit properties of the ground states.

Our main results are the upper and lower bounds on the gap in the
spectrum of the Hamiltonian in the thermodynamic limit in the
aforementioned four superselection sectors. Each of these sectors
corresponds to a different representation of the observable algebra of
the system. In these representations the Heisenberg dynamics of the
model is generated by a densely defined self-adjoint, non-negative
definite operator $H$. Theorem \ref{thm:main} below refers to the gap
above zero in the spectrum of this operator. Alternatively, we can
define the gap of the model with respect to an infinite volume ground
state $\langle \,\cdot\, \rangle$ as the largest constant $\gamma\geq
0$ such that for all {\it local\/} observables $X$
\be
\langle X^* H^3 X\rangle=\langle X^* [H^3,X]\rangle
\geq \gamma\langle X^* [H^2,X]\rangle =\gamma\langle X^* H^2 X\rangle
\label{defgap1}\ee
By a local observable $X$ we simply mean here a polynomial of a finite
number of spin matrices. All local observables obtained from spin
matrices $S^\alpha_x$ with site index $x$ in a given finite subset
$\Lambda$ of the chain form an algebra which is denoted by
$\A_\Lambda$. The inequality \eq{defgap1} expresses the property that
the Hamiltonian is larger than $\gamma$ on its range, i.e., on the
othogonal complement of the space of ground states. If the ground
state is non-degenerate in the representation under consideration, the
following equivalent inequality is more customary:
\be
\langle X^* [H,X]\rangle\geq \gamma(\langle X^*X\rangle
-\vert\langle X\rangle\vert^2)
\label{defgap2}\ee
In both \eq{defgap1} and \eq{defgap2} the commutator has to be
interpreted as the limit of successive commutators
\be
[H^p,X]=\lim_{\Lambda_p\to \infty} \cdots \lim_{\Lambda_1\to \infty}
[H_{\Lambda_p} , [H_{\Lambda_{p-1}},\cdots[H_{\Lambda_1},X]\cdots ]]
\ee
which, due to the fact that the interaction is of finite range,
is a local observable for all local $X$.

\begin{Thm}\label{thm:main}
In each of the sectors described above as {\em up,
down, kink\/}, and {\em antikink\/},
the infinite volume gap $\gamma$ satisfies
\be
\Gammalower(\Delta)\leq \gamma\leq \Gammaupper(\Delta)
\label{boundsgamma}\ee
for $\Delta >1$ and where
\bea
\Gammaupper(\Delta) &=& 1-\Delta^{-1}\label{Gammaupper}\\
\Gammalower(\Delta) &=&
\left(1-\sqrt{1-\sqrt{1-\frac{1}{\Delta^2}}}\;\right)^2
\label{Gammalower}\eea
For $\Delta =1$, the difference $\gamma_L$ between the
lowest and second-lowest eigenvalue of the Hamiltonian $H_L$ of a
finite chain of length $L$, $L\geq 2$, satisfies
\be
\frac{1}{4L^2}\leq (1-\sqrt{1-\frac{1}{L-1}})^2\leq \gamma_L
\leq 4\pi^2(\frac{1}{L^2}+\frac{4}{L^3})
\label{isotropicgap}
\ee
\end{Thm}

Two other parametrizations common in the literature are given by
$\Delta=\cosh\lambda=(\mu+\mu^{-1})/2$. The range $1\leq \Delta \leq
+\infty$ corresponds to $0\leq \lambda \leq +\infty$ and $1\geq \mu
\geq 0$. The parameter $\mu$ is often denoted by $q$. The most
common Hamiltonian is $\Delta$ times $H^{XXZ}$ defined in \eq{ham+bc}
(up to the boundary terms). In terms of the parameter $\lambda$ the
bounds of Theorem \ref{thm:main} are
\bea
\Delta\Gammaupper &=& \cosh\lambda -1
\label{Gammaupperlambda}\\
\Delta\Gammalower &=&
\cosh\lambda\left(1-\sqrt{1-\tanh\lambda}\;\right)^2
\label{Gammalowerlambda}
\eea
Near the isotropic ferromagnet ($\Delta =1$) the upper and lower
bound both behave linearly, with a slope $2$ and $1$ respectively.
This fixes the critical exponent $\alpha$ which governs the behavior
of the gap (as well as the low-temperature behaviour of the specific
heat) near $\Delta=1$ to be equal to $1$, which is in agreement
with the exact result of Johnson, Krinsky and McCoy \cite{JKMcC}.
Near $\Delta=+\infty$ the lower bound behaves
as $1-\sqrt{2}/\Delta$.

Figure 1 gives an idea of the difference between the bounds
$\Gammalower$ and $\Gammaupper$. $\Gammaupper$, here derived by a
variational argument in Section \ref{sec:infinite}, is the exact
solution given by the Bethe Ansatz \cite{JMcC,Klu,AKS}. Also for
the finite volume estimates at $\Delta =1$, there are exact
expressions for the coefficient of $1/L^2$ \cite{ABB}.

The advantages of the approach in this paper are: 1) The method can be
applied also to Hamiltonians that cannot be explicitly diagonalized;
2) As the the treatment is short, transparent, and completely
rigorous, it should also deliver a better insight in the determining
properties of the gap; 3) It is possible to obtain useful bounds down
to the critical point (here $\Delta=1$), whereas this is usually not
possible with other rigorous methods such as, e.g., the polymer
expansion technique of \cite{KT}.


\newsection{The ground states of the XXZ chain}\label{sec:ground}

Only these aspects of the ground states of the XXZ chain that have
direct relevance to our estimates and understanding of the spectral
gap of the model will be presented here. A more detailed  analysis can
be found in \cite{GW}, and various aspect of the ground states have
been discussed in the literature (see e.g. \cite{ASW} and the
references therein). It  should be mentioned that a full analysis of
the ground state problem for the infinite chain has not been achieved
yet. Below we give a clear description of what is believed to be the
complete set of ground states for the infinite chain. I am not aware
of a rigorous proof that this is indeed the case. Loosely speaking one
would obtain a description of the complete set of ground states by
studying the thermodynamic limit with arbitrary boundary conditions.
The difficulty is that a simple description of a sufficiently large
class of boundary conditions is not available. Fortunately the
statements in this article do not depend on the completeness of the
set of ground states considered.

For the study of the finite chains in this section we shall employ the
special boundary conditions introduced in \eq{ham+bc}.  This choice of
boundary conditions simplifies the study of the thermodynamic limit.
It is also convenient to add a constant to the Hamiltonian to make the
ground state energy vanish.  This way, using the parameter $\mu$, we
arrive at the Hamiltonian,
\begin{equation}
H^\mu_L=H^{XXZ}_L +(L-1)/4=\sum_{x=1}^{L-1}h^\mu_{x,x+1}
\label{hammu}
\end{equation}
where $h^\mu_{x,x+1}$ is the orthogonal projection on the vector
\begin{equation}
\xi_\mu=\frac{1}{1+\mu^2}(\mu\ket{\up \down }-\ket{\down \up })
\label{ximu}
\end{equation}
In terms of the spin matrices $h^\mu_{1,2}$ is
\be
h^\mu_{1,2}=- \Delta^{-1}(S^1_1 S^1_2 + S^2_1 S^2_2) - S^3_1 S^3_2
+{1\over 4} +A(\Delta)(S^3_2 - S^3_1)
\ee
with $A$ defined following \eq{ham+bc}. From the definition of
$\xi_\mu$ it is obvious that $h^\mu_{x,x+1}\ket{\up\cdots\up}=0$ for all
$x=1,\ldots,L-1$. As $H^\mu_L$ is the sum of the $h^\mu_{x,x+1}$,
which are positive, this implies that the ground state  energy of
$H^\mu_L$ is zero and that $\ket{\up\cdots\up}$ is a ground state. For
all $0\leq \mu\leq 1$, the ground state space ($\equiv\ker H^\mu_L$)
is $L+1$-dimensional. An explicit description of the $L+1$ ground
states can be given in several ways.

For all $\Delta \geq 1$ the uniform states $\ket{\up\cdots\up}$ and
$\ket{\down\cdots\down}$ are ground states of the XXZ chain.
If $\Delta =1$ the $L+1$-dimensional ground state space is the
spin $L/2$ representation of SU(2). For all $\Delta >1$ and
$A=\frac{1}{2}\sqrt{1-1/\Delta^2}$, the non-uniform ground states can
be thought of as kink states, which are roughly described as the Ising
kinks plus quantum fluctuations. In this picture the degeneracy
corresponds to the possible locations of the  kink. For
$A=-\frac{1}{2}\sqrt{1-1/\Delta^2}$ the kinks have to be replaced by
antikinks, i.e., the roles of $\up$ and $\down$ spins have to be
interchanged (or, equivalently, one can interchange left and right).
We refer to \cite{ASW} and \cite{GW} for more details and explicit
expressions.

In the thermodynamic limit the boundary terms disappear to infinity and
the left-right symmetry of the model, broken by the particular
boundary terms we have introduced, must be restored. It is therefore
obvious that both the kink and antikink states appear as infinite
volume ground states of the model.

For our purposes the most convenient way to describe the space of
ground states of a chain of length $L$ is to introduce deformed
raising and lowering operators which, together with the third
component of the spin, generate the algebra (quantum group) of
SU$_\mu$(2). The concrete representation of SU$_\mu$(2) is {\it not\/}
left-right symmetric, and is different for the boundary terms that
produce kink and antikink ground states. In fact the two mutually
non-commuting representations of SU$_\mu$(2) together generate the
infinite-dimensional quantum affine symmetry algebra $\widehat{sl}(2)$
that lies at the basis of the integrability of the model (see e.g.
\cite{DFJMN}). We should stress, however, that a rigorous formulation
of this infinite dimensional symmetry of the XXZ chain, has not yet
been obtained. We will not use it here.

In our computations we will not need anything beyond some basic facts
of the representation theory of SU$_\mu$(2). We therefore restrict the
discussion of the quantum group symmetry of the XXZ model to the bare
minimum. One can think of the quantum group symmetry as a systematic way
to construct operators that commute with the Hamiltonians $H^\mu_L$. The
parallellism with the usual arguments in the ``theory of angular
momentum'' in quantum mechanics (representations of  SU(2)) is so
perfect that the reader will hardly notice the difference.

For $0<\mu<1$ define the $2\times 2$ matrix $t$ by
\be
t=\mu^{-2S^3}
\label{deft}
\ee
and define as usual $S^\pm=S^1\pm i S^2$. It is trivial to check that
$S^\pm$ and $t$ satisfy the following commutation relations
\alpheqn\bea
tS^\pm&=&\mu^{\mp2}S^\pm t \label{sumua}\\
{[ S^+ , S^- ] }&=&\frac{\displaystyle t-t^{-1}}{\displaystyle
\mu^{-1}-\mu}
=2S^3\label{sumub}
\eea\reseteqn
They are just the SU(2) commutation relations in a disguised form. The
remarkable fact is that there is a simple definition of the tensor
product (coproduct of the quantum group or pseudogroup \cite{Dri,Wor})
of any two representations of the commutation relations
\eq{sumua}--\eq{sumub}, yielding a new representation. Here we only
need the total-spin operators for a chain of $L$ spins, which are
given by
\alpheqn\bea
S^3_{[1,L]}&=&\sum_{x=1}^L \idty_1\otimes\cdots\otimes
S^3_x\otimes\idty_{x+1}\otimes\cdots\idty_L\label{spinmua}\\
S^+_{[1,L]}&=&\sum_{x=1}^L t_1\otimes\cdots\otimes t_{x-1}\otimes
S^+_x\otimes\idty_{x+1}\otimes\cdots\idty_L\label{spinmub}\\
S^-_{[1,L]}&=&\sum_{x=1}^L \idty_1\otimes\cdots\otimes
S^-_x\otimes t^{-1}_{x+1}\otimes\cdots t^{-1}_L\label{spinmuc}
\eea\reseteqn
where we used an index to identify the sites on which the tensor
factors act. Note that, for $L\geq 2$, the operators $S^\pm_{[1,L]}$
depend on $\mu$ through $t$. One can easily check that the total
``spin'' operators as defined in \eq{spinmua}--\eq{spinmuc} commute
with the interaction terms $h^\mu_{x,x+1}$ and hence with the
Hamiltonian $H^\mu_L$ itself.

\newsection{Estimate of the mass gap for finite chains}\label{sec:gap}

We begin this section with the derivation of a simple lower bound for
the spectral gap of finite chains for Hamiltonians that share some of
the basic properties observed in the XXZ chain (Theorem
\ref{thm:generalestimate}). A more general version of this estimate
was given in \cite{GVBS} where it was used to prove the existence of a
spectral gap in arbitrary Generalized Valence Bond Solid chains with a
finite number of ground states. As a strategy for obtaining lower
bounds for the spectral gap of the generator of a spin dynamics, the
method of proof is inspired by the work of Lu and Yau \cite{LY} on the
gap in the spectrum of the Glauber and Kawasaki dynamics of the Ising
model. The ingredients that go into the estimate are not very
different from the ones in \cite{FCS} and in fact similar elements
underly the arguments in \cite{AKLT,Zeg,MO}.

Here, we restrict ourselves to the simplest form of this estimate,
which is sufficient for the application to the XXZ chain.

Consider an arbitrary spin chain of $L$ sites and with Hilbert space
$\Hs_L=\bigotimes_{x=1}^L (\Cx^d)_x$, where again we use the index $x$
to associate the tensor factors with the sites in the chain. We assume
that the Hamiltonian is of the following form:
\be
H_L=\sum_{x=1}^{L-1} h_{x,x+1}
\label{HL}
\ee
where $h_{x,x+1}$ is a translation of $h_{1,2}$, acting non-trivially
only at the nearest neighbour pair $\{x,x+1\}$. Assume furthermore
that $h_{1,2}\geq 0$ and that $\ker H_L \neq\{0\}$. We will denote by
$\gamma_2$ the smallest nonzero eigenvalue of $h_{1,2}$, i.e., the gap
of $H_2$. It is obvious that
\be
\ker H_L =\bigcap_{x=1}^{L-1} \ker h_{x,x+1}
\label{kerH}
\ee
For an arbitrary subset
$\Lambda$ let $G_\Lambda$ be the orthogonal projection onto
\be
\ker \sum_{x, \{x,x+1\}\subset\Lambda} h_{x,x+1}
\label{GLambda}
\ee
For intervals $[a,b]$, $1\leq a < b\leq L$, $G_{[a,b]}$ is the
orthogonal projection onto the zero eigenvectors of $\sum_{x=a}^{b-1}
h_{x,x+1}$, and $G_{\{x\}}=\idty$ for all $x$. From these definitions
it immediately follows that the orthogonal projections $G_\Lambda$
satisfy the following properties:
\alpheqn\bea
G_{\Lambda_2}G_{\Lambda_1} &=& G_{\Lambda_1}G_{\Lambda_2}
= G_{\Lambda_2} \mbox{ if } \Lambda_1 \subset\Lambda_2
\label{GLambdaa}\\
G_{\Lambda_1}G_{\Lambda_2} &=& G_{\Lambda_2}G_{\Lambda_1}
\mbox{ if } \Lambda_1 \cap \Lambda_2 =\emptyset
\label{GLambdab}\\
h_{x,x+1} &\geq& \gamma_2(\idty-G_{[x,x+1]})
\label{GLambdac}
\eea\reseteqn
Define operators $E_n$, $1\leq n\leq L$, on $\Hs_L$ by
\be
E_n=\cases{\idty-G_{[1,2]}       & if $n=1$\cr
           G_{[1,n]}-G_{[1,n+1]} & if $2\leq n \leq L-1$\cr
            G_{[1,L]}            & if $n=L$\cr }
\label{defEn}\ee
One can then easily verify, using the properties
\eq{GLambdaa}-\eq{GLambdac}, that $\{E_n \mid1\leq n\leq L\}$
is a family of mutually orthogonal projections summing up to $\idty$,
i.e.:
\be
E_n^*=E_n,\qquad E_n  E_m=\delta_{m,n} E_n,
\qquad \sum_{n=1}^L E_n =\idty
\label{resolution}\ee

The preceding paragraph applies directly to the XXZ chain. Next, we
make a non-trivial assumption which we will verify for the XXZ chain
later.

\begin{Ass}\label{thm:ass}
There exists a constant $\epsilon$, $0\leq \epsilon
< 1/\sqrt{2}$, such that for all $1\leq n\leq L-1$
\be
E_n G_{[n,n+1]} E_n \leq \epsilon^2 E_n
\label{assa}\ee
or, equivalently,
\be
\Vert G_{[n,n+1]} E_n\Vert \leq \epsilon
\label{assb}\ee
\end{Ass}

Note that, due to \eq{GLambdaa}, $G_{[n,n+1]} E_n= G_{[n,n+1]}
G_{[1,n]} - G_{[1,n+1]}$. This relates Assumption \ref{thm:ass}
with Lemma 6.2 in \cite{FCS}, where an estimate for $\Vert G_{[n,n+1]}
G_{[1,n]} - G_{[1,n+1]}\Vert$ is given for general Valence Bond Solid
chains with a unique infinite volume ground state. The same
observation also implies that $[G_{[n,n+1]}, G_{[1,n]}]= [G_{[n,
n+1]}, E_n]$, which, if \eq{assb} holds, is bounded above in norm by
$2\epsilon$.

The next theorem is a special case of Theorem 2.1 in \cite{GVBS}. Just
like Theorem 6.4 in \cite{FCS} it provides a lower bound on the gap of
the finite volume Hamiltonians, but it achieves this in a slightly
more efficient way. We will repeat the proof for the particular case
stated  here, because it is simple, short, and instructive.

\begin{Thm}\label{thm:generalestimate}
With the definitions of above and under Assumption \ref{thm:ass}
the following estimate holds for all $\psi$ satisfying
$G_{[1,L]} \psi=0$, i.e.,
$\psi$ that are orthogonal to the space of ground states of
$H_L$ :
\be
\langle \psi\mid H_L\psi\rangle
\geq \gamma_2 (1-\sqrt{2}\epsilon)^2\Vert\psi\Vert^2
\label{gapestimate}\ee
i.e., the spectrum of $H_L$ has a gap of at least
$\gamma_2 (1-\sqrt{2}\epsilon)^2$ above the lowest eigenvalue, which is
$0$.
\end{Thm}

\begin{proof}
{}From the properties \eq{resolution} of the $E_n$ and the
assumption that $G_{[1,L]}\psi=0$, it immediately follows that
\be
\Vert\psi\Vert^2=\sum_{n=1}^{L-1}\Vert E_n\psi\Vert^2
\label{resolutionnormpsi}\ee
One can estimate $\Vert E_n\psi\Vert^2$ in terms of $\langle
\psi\mid h_{n,n+1}\psi\rangle$ as follows. First insert $G_{[n,n+1]}$
and the resolution $\{ E_m\}$:
\be
\Vert E_n\psi\Vert^2
=\langle\psi\mid(\idty-G_{[n,n+1]})E_n\psi\rangle
+ \langle\psi\mid \sum_{m=1}^{L-1} E_m G_{[n,n+1]}E_n\psi\rangle
\label{Enpsi}\ee
Using \eq{GLambdaa} and  \eq{GLambdab} one easily veryfies that $E_m$
commutes with $G_{n,n+1}$ if either $m\leq n-2$ or $m\geq n+1$. In
these cases $E_m G_{[n,n+1]}E_n= G_{[n,n+1]} E_m E_n=0$, because the
$E_n$ form an orthogonal family. By this observation we obtain the
following estimate. For any choice of constants $c_1,c_2>0$:
\bea
\Vert E_n\psi\Vert^2 &=&
\langle\psi\mid(\idty-G_{[n,n+1]})E_n\psi\rangle
+ \langle (E_{n-1}+E_n) \psi\mid G_{[n,n+1]}E_n\psi\rangle
\nonumber\\
&\leq& {1\over 2 c_1} \langle\psi\mid(\idty-G_{[n,n+1]})\psi\rangle +
{c_1\over 2}\langle\psi \mid E_n\psi\rangle
\label{Enpsi2}\\
&&\quad  +{1\over 2c_2} \langle\psi\mid E_n G_{[n,n+1]}
E_n\psi\rangle+{c_2\over 2}\langle\psi \mid (E_{n-1}+E_n)^2\psi\rangle
\nonumber\eea
where we have applied the inequality
$$
\vert\langle\phi_1\mid\phi_2
\rangle\vert\leq {1\over 2 c}\Vert \phi_1\Vert^2 +{c\over
2}\Vert\phi_2\Vert^2\quad ,
$$
for any $c>0$, to both terms of \eq{Enpsi}. The first term in the right
side of inequality \eq{Enpsi2} can be estimated with the interaction
using \eq{GLambdac}. The third term can be estimated with \eq{assa}.
It then follows that
$$
(2-c_1-{\epsilon^2\over c_2})\Vert E_n\psi \Vert^2
-c_2 \Vert (E_{n-1}+E_n)\psi\Vert^2
\leq {1\over c_1 \gamma_2}\langle\psi
\mid h_{n,n+1} \psi \rangle
$$
The term containing $E_{n-1}$ is absent for $n=1$, and
$E_n\psi=0$ if $n=L$.
We now sum over $n$ and use \eq{resolutionnormpsi} to obtain
$$
(2-c_1-{\epsilon^2\over c_2}-2c_2)\Vert\psi\Vert^2
\leq {1 \over c_1 \gamma_{2}}\langle\psi
\mid H_L \psi \rangle
$$
Finally put $c_1=1-\epsilon\sqrt{2}$ and
$c_2=\epsilon/\sqrt{2}$ and
one obtains the estimate \eq{gapestimate} stated in the theorem.
\end{proof}

We now return to the XXZ chain and prove a lower bound on the gap of
the finite volume Hamiltonians \eq{ham+bc} as an application of
Theorem \ref{thm:generalestimate}.

\begin{Prop}\label{thm:Gammalower}
For a spin 1/2 chain of length $L$, the Hamiltonians
\be
H^{XXZ}_L=A(S_L^3-S_1^3)
-\sum_{x=1}^{L-1} \{\frac{1}{\Delta}(S^1_xS^1_{x+1}+S^2_xS^2_{x+1})
+S^3_xS^3_{x+1}-{1\over 4}\}
\label{haml}\ee
with $A=\pm {1\over 2}\sqrt{1-1/\Delta^2}$, $\Delta\geq 1$,
have an $L+1$-fold degenerate ground state with eigenvalue $0$ and
their next largest eigenvalue, $\gamma_L$, satisfies
\be
\gamma_L\geq (1-\sqrt{\frac{2\mu}{\mu+\mu^{-1}}})^2
= (1-\sqrt{1-\tanh \lambda})^2 =\Gammalower(\Delta)
\ee
where $\Gammalower(\Delta)$ is defined in \eq{Gammalower}.
If $\Delta =1$ one has the lower bound
\be
\gamma_L\geq (1-\sqrt{1-\frac{1}{L-1}})^2
\geq \frac{1}{4L^2}
\ee
\end{Prop}

\begin{proof}
Due to the reflection (left-right) symmetry of the interaction it is
clearly sufficient to consider one sign of $A$ in \eq{haml}, say
$A\geq 0$. The proof consists in giving a constant $\epsilon$ for
which Assumption \ref{thm:ass} holds. In fact, for the model under
consideration one can simply compute the quantity $\Vert G_{[n,n+1]}
E_n\Vert$. The spaces on which the $G_{[a,b]}$ project are described
explicitly in Section \ref{sec:ground}. Here, we consider  all
operators as acting on $\Hs_L$. The quantity $\Vert G_{[n,n+1]}
E_n\Vert$ is equal to $C_n$ defined for $n\geq 1$ by
\be
C_n=\sup_{0\neq\psi\in\Hs_{n+1}\atop E_n \psi=\psi}
\frac{\Vert G_{[n,n+1]}\psi\Vert}{\Vert\psi\Vert}
\label{Cn}\ee
It follows from \eq{GLambdaa} that $\Vert E_n\psi\Vert^2
=\Vert G_{[1,n]}\psi\Vert^2 - \Vert G_{[1,n+1]}\psi\Vert^2$.
Therefore, the requirement $E_n\psi=\psi$ can be expressed as
\be
G_{[1,n]}\psi=\psi,\qquad G_{[1,n+1]}\psi=0
\label{supconditions}\ee
As long as $1\leq n\leq L-1$, $C_n$  does not depend on $L$.

The representation theory of SU$_\mu$(2) is isomorphic to the one of
SU(2) \cite{Wor2}. The irreducible representations can be labeled by
the half-integers $s=0,1/2,1,3/2,\ldots$, and we denote them by
$D^{(s)}$. We will use a subscript to indicate the set of sites on
which a particular representation acts. $D^{(s)}$ is realized on
$\Hs_{2s} \cong (\bigotimes\Cx^2)^{2s}$ as the space of ground states
of $H^\mu_{2s}$. These elementary facts are sufficient to determine
the vectors $\psi$ satisfying \eq{supconditions}. Indeed,
$G_{[1,n]}\psi=\psi$ implies that $\psi\in D^{(n/2)}_{[1,n]}\otimes
D^{(1/2)}_{\{n+1\}}\subset \Hs_{n+1}$. As $D^{(n/2)}\otimes
D^{(1/2)}\cong D^{((n-1)/2)}\oplus D^{((n+1)/2)}$, $G_{[1,n+1]}\psi=0$
implies that $\psi\in D^{((n-1)/2)}$.

The ratio of norms in \eq{Cn} is invariant under the action of
SU$_\mu$(2), and is therefore the same for all $\psi\in
D^{((n-1)/2)}$. One shows this just as one would for a group
representation. For completeness we include a detailed argument here.
$G_{[n,n+1]}=\idty-\ket{\xi_\mu}\bra{\xi_\mu}$ commutes with
$S^3_{[1,n+1]}$. Therefore, it is sufficient to consider $\psi$ that
are eigenvectors of $S^3_{[1,n+1]}$. Starting form one such
eigenvector we can generate all others by application  of
$S^\pm_{[1,n+1]}$. Assuming that $S^\pm_{[1,n+1]}\psi\neq 0$ and using
the fact that $G_{[n,n+1]}$ also commutes with $S^\pm_{[1,n+1]}$ we
find
\be
\frac{\Vert G_{[n,n+1]}S^\pm_{[1,n+1]}\psi\Vert^2}{\Vert
S^\pm_{[1,n+1]}\psi\Vert^2}=\frac{\langle \psi \mid
G_{[n,n+1]}(S^\pm_{[1,n+1]})^*S^\pm_{[1,n+1]}\psi\rangle}{\langle
\psi\mid(S^\pm_{[1,n+1]})^*S^\pm_{[1,n+1]}\psi\rangle}
\label{same}\ee
As $(S^+_{[1,n+1]})^*=\mu^{-1} t_{[1,n+1]}S^-_{[1,n+1]}$
(see \eq{spinmua} and \eq{spinmub}),
$(S^\pm_{[1,n+1]})^*S^\pm_{[1,n+1]}\psi$ is an eigenvector of
$S^3_{[1,n+1]}$ with the same eigenvalue as $\psi$. This vector
also belongs to the same irreducible representation and hence it
must be proportional to $\psi$. It follows that the ratio in
\eq{same} is the same as for $\psi$.

The computation of $C_n$ is now straightforward, for we have to
consider just one vector $0\neq \psi\in D^{((n-1)/2)}$,
e.g., one with $S^3_{[1,n+1]}=(n-1)/2$. Any such vector is of the form
\be
\psi=\sum_{x=1}^{n+1} a_x D_x
\ee
where $D_x$ denotes the ususal basis vector with all spins up
except at the site $x$ where the spin is down. Up to normalization
the coefficients $a_x$ are uniquely determined by the conditions
\eq{supconditions}. A possible choice is
\be
\psi= \frac{\mu^n-\mu^{-n}}{\mu-\mu^{-1}}D_{n+1}
-\sum_{x=1}^n \mu^{-x} D_x
\ee
A short computation shows that
\bea
\Vert\psi\Vert^2 &=& \frac{(\mu^n-\mu^{-n})(\mu^n-\mu^{-n-2})}{%
(\mu-\mu^{-1})^2}\\
\Vert G_{[n,n+1]}\psi\Vert^2 &=&
\frac{(\mu^{n-1}-\mu^{-n+1})(\mu^n-\mu^{-n-2})}{
(\mu-\mu^{-1})^2(\mu+\mu^{-1})}
\eea
and therefore
\be
C_n=\frac{\sqrt{\mu(1-\mu^{2n-2})}}{\sqrt{(\mu+\mu^{-1})(1-\mu^{2n})}}
\ee
which is increasing in $n$ and $C_n< 1/\sqrt{2}$ for all $n\geq1$.
We conclude that the assumptions of Theorem \ref{thm:generalestimate}
are satisfied with
\be
\epsilon=C_{L-1}< \sqrt{\frac{\mu}{\mu+\mu^{-1}}}
\ee
This proves the theorem for $0<\mu<1$. For $\mu=1$ one has to
consider the limit
\be
\epsilon=\lim_{\mu\up 1} C_{L-1}=\frac{1}{\sqrt{2}}
\sqrt{1-\frac{1}{L-1}}
\ee
which is straightforward to compute.
\end{proof}

\newsection{The infinite chain}\label{sec:infinite}

In order to be able to prove rigorous statements about the spectrum of
the infinite chain we need to introduce the mathematical objects that
define the infinite system. Although all interesting properties of the
infinite chain can be expressed as results for limits of quantities
defined for finite chains, the reverse is not true. Not all limits of
finite chain quantities give interesting or even sensible statements
about the infinite chain. By using a clean definition of the infinite
system we will have no difficulty in sorting out the relevant
statements about the thermodynamic limit of the XXZ chain.

Let the symbols $\up\up,\up\down,\down\up,\down\down$ denote the four
superselection sectors of the infinite XXZ chain with $\Delta>1$,
corresponding to {\it up, kinks, antikinks\/}, and {\it down\/}
respectively. We can describe the GNS Hilbert spaces \cite{BraRo} of
these four superselection sectors as the so-called incomplete tensor
products \cite{Gui} $\Hs_{\alpha\beta}$, for $\alpha$ and $\beta=\up$
or $\down$, defined by
\begin{equation}
\Hs_{\alpha\beta}=\overline{\bigcup_\Lambda
\left(\bigotimes_{x\in\Lambda}
\Cx^2\otimes\bigotimes_{y\in\Lambda^c}\Omega_{\alpha\beta}(y)\right)}
\label{Hab}
\end{equation}
where
\begin{equation}
\Omega_{\alpha\beta}(y)=
\cases{\ket{\alpha}&if $y\leq 0$\cr
       \ket{\beta}&if $y> 0$\cr}
\label{Oaby}
\end{equation}
We also define the vectors $\Omega_{\alpha\beta}$ as the infinite
product vectors
\begin{equation}
\Omega_{\alpha\beta}=\bigotimes_{y\in\Ir}\Omega_{\alpha\beta}(y)
\in\Hs_{\alpha\beta}
\label{Oab}
\end{equation}
Let $\A_\Lambda$ denote the local observables acting nontrivially only
on the sites in the finite set $\Lambda$. Local observables
$X\in\A_\Lambda$ act on $\Hs_{\alpha\beta}$ in the obvious way, e.g.,
the spin matrices at the site $x$ act on the $x^{th}$ factor of the
tensor product \eq{Hab}. From the definitions above it is clear that
vectors $\psi$ of the form
\begin{equation}
\psi=X\Omega_{\alpha\beta},\qquad X\in \bigcup_\Lambda \A_\Lambda
\label{psi}
\end{equation}
form a dense subspace of $\Hs_{\alpha\beta}$. Note that if
$\alpha\neq\beta$, $\Omega_{\alpha\beta}$ is {\it not\/} the GNS
vector representing one of the kink (or antikink) ground states.

The mass gap of the infinite chain is a property which is defined with
respect to a particular ground state of the infinite system or, more
precisely, with respect to a superselection sector.  The Hamiltonian
is represented on $\Hs_{\alpha\beta}$ as the generator
$H_{\alpha\beta}$ of the Heisenberg dynamics of observables acting on
$\Hs_{\alpha\beta}$.  The dense subspace of the vectors $\psi$ defined
in \eq{psi} is in the domain of $H_{\alpha\beta}$, and the selfadjoint
operator $H_{\alpha\beta}$ is uniquely determined by the requirement
\begin{equation}
H_{\alpha\beta}X\Omega=\lim_{\Lambda\to\Ir}[H^{XXZ}_\Lambda,X]
\Omega_{\alpha\beta}
\label{hamab}
\end{equation}
We remark that $H_{\alpha\beta}$ does not depend on boundary terms such
as $A(S^3_a-S^3_b)$ added to the XXZ Hamiltonian for finite chains. It
is well-known \cite{BraRo} that $H_{\alpha\beta}$ is a positive
operator in general and in the present case this could not be more
clear. An explicit formula for $H_{\alpha\beta}$ is
\begin{equation}
H_{\alpha\beta}X\Omega_{\alpha\beta}= \sum_{\{x,x+1\}\cap \Lambda
\neq 0} h^{\alpha\beta}_{x,x+1} X\Omega_{\alpha\beta}
\label{hamloc}
\end{equation}
where $h^{\alpha\beta}_{x,x+1}$ can be taken to be $h^\mu_{x,x+1}$ if
$\alpha\beta=\up\up, \down\down$, or $\up\down$. If $\alpha\beta
=\down\up$ the sign of the boundary term has to be reversed. This is
equivalent to replacing $\mu$ by $\mu^{-1}$.

The mass gap $\gamma_{\alpha\beta}$ is then just the gap  above $0$ in
the spectrum of $\H_{\alpha\beta}$. A formula for
$\gamma_{\alpha\beta}$ is
\be
\gamma_{\alpha\beta}=\inf_{0\neq\psi\perp\ker
H_{\alpha\beta}\atop \psi\in\dom H_{\alpha\beta}}\frac{\langle\psi\mid
H_{\alpha\beta}\psi\rangle}{\langle\psi\mid\psi\rangle}
\label{defGamma}
\ee
As $(\ker H_{\alpha\beta})^\perp =\ran H_{\alpha\beta}$ the infimum in
\eq{defGamma} can be taken over vectors of the form
$H_{\alpha\beta}\psi$, and it suffices to take $\psi$ of the form
\eq{psi} because they are a core for $H_{\alpha\beta}$.

There is no a priori reason why the spectrum of $H_{\alpha\beta}$
should be independent of the superselection sector, i.e. independent
of $\alpha\beta$. We already know that the multiplicity of the lowest
eigenvalue is different: it is 1 for $H_{\up\up}$ and $H_{\down\down}$
and infinite for $H_{\up\down}$ and $H_{\down\up}$. Therefore, a
priori, we should not expect $\gamma_{\alpha\beta}$ to be independent
of $\alpha\beta$. One can easily convince oneself, however, that
$\gamma_{\alpha\beta}= \gamma_{\beta\alpha}$ and that
$\gamma_{\up\up}=\gamma_{\down\down}$. From a simple argument given in
Section \ref{sec:upper} it follows that
$\gamma_{\alpha\beta}\leq\gamma_{\up\up}$. The upper and lower bounds
that we will derive here are independent of $\alpha\beta$.

\subsection{Proof of the lower bound of Theorem \protect\ref{thm:main}}
\label{sec:lower}

In order to prove the lower bound of \eq{boundsgamma} we simply have
to show that the lower bound on the finite volume gap obtained in
Section \ref{sec:gap} remains valid in the thermodynamic limit,
irrespective of the particular zero energy ground state that we are
considering. It is important that the finite volume gap estimates
were obtained for the ``correct'' boundary conditions of \eq{ham+bc}.
More explicitly we show that for any choice of $\alpha\beta$
and all local observabels $X$ the following inequality holds:
\be
\langle\Omega_{\alpha\beta}\mid X^* H_{\alpha\beta}^3 X
\Omega_{\alpha\beta}\rangle
\geq \Gammalower(\Delta)\langle\Omega_{\alpha\beta}\mid X^*
H_{\alpha\beta}^2 X\Omega_{\alpha\beta}\rangle
\label{lowerbound}
\ee
If $\alpha=\beta$, $\Omega_{\alpha\beta}$ is the vector representing
the unique ground state of $H_{\alpha\beta}$. If $\alpha\neq\beta$,
$\Omega_{\alpha\beta}$ is not a ground state itself (except in the
Ising limit $\Delta=\infty$). But all the kink (if
$\alpha\beta=\up\down$) or antikink states (if $\alpha\beta=\down\up$)
are represented as vectors in the Hilbertspace $\Hs_{\alpha\beta}$
defined by \eq{Hab}, and together these vectors span $\ker
H_{\alpha\beta}$. A proof of these statements can be found in
\cite{GW}.

The inequality \eq{lowerbound} follows from Proposition
\ref{thm:Gammalower} when one observes that for $X\in\A_\Lambda$
\be
\langle\Omega_{\alpha\beta}\mid X^* H_{\alpha\beta}^3 X
\Omega_{\alpha\beta}\rangle
= \langle\Omega_{\alpha\beta}\mid X^* (H^\mu_{\Lambda\pm 3})^3 X
\Omega_{\alpha\beta}\rangle
\label{local}
\ee
Obviously, $X^* (H^\mu_{\Lambda\pm 3})^3 X\in \A_{\Lambda\pm 3}$.
Therefore the expectation value in the right side of \eq{local} can be
computed in the density matrix $\rho_{\Lambda\pm 3}$ which describes
the state $\Omega_{\alpha\beta}$ in the finite volume $\Lambda \pm 3$.
The same is true for the right side of \eq{lowerbound}. We conclude
that it is sufficient to ascertain that
\be
\Tr \rho_{\Lambda\pm 3} X^* (H^\mu_{\Lambda\pm 3})^3 X
\geq \Gammalower(\Delta)\Tr \rho_{\Lambda\pm 3} X^* (H^\mu_{\Lambda
\pm 3})^2 X
\ee
which immediately follows from the finite volume gap estimate of
Proposition \ref{thm:Gammalower}.
\endproof

\subsection{Proof of the upper bound of Theorem \protect\ref{thm:main}}
\label{sec:upper}

First we argue that it suffices to prove the upper bound for
$H_{\up\up}$. It is obvious that the gap of $H_{\down\down}$ will
satisfy the same bound. For the gap of the model in the kink and
antikink sectors we have an inequality which can be derived as
follows. The translation invariant ground states can be  obtained as
weak limits of the kink or antikink states by letting the position of
the kink (or antikink) tend to $\pm\infty$. We then have
\bea
\lefteqn{\inf_{\Lambda,X\in\A_\Lambda}
\frac{\langle\Omega_{\alpha\beta}\mid X^* (H^\mu_{\Lambda\pm 3})^3 X
\Omega_{\alpha\beta}\rangle}{\langle\Omega_{\alpha\beta}\mid X^*
(H^\mu_{\Lambda\pm 3})^2 X \Omega_{\alpha\beta}\rangle}}\\
&\leq \displaystyle \inf_{\Lambda,X\in\A_\Lambda} \lim_{n\to\pm\infty}
\frac{\langle\Omega_{\alpha\beta}\mid\tau_n(X^*
(H^\mu_{\Lambda\pm 3})^3 X)
\Omega_{\alpha\beta}\rangle}{\langle\Omega_{\alpha\beta}\mid \tau_n(X^*
(H^\mu_{\Lambda\pm 3})^2 X)\Omega_{\alpha\beta}\rangle}\\
&= \displaystyle\inf_{\Lambda,X\in\A_\Lambda}
\frac{\langle\Omega_{\up\up}\mid X^* (H^\mu_{\Lambda\pm 3})^3 X
\Omega_{\up\up}\rangle}{\langle\Omega_{\up\up}\mid X^*
(H^\mu_{\Lambda\pm 3})^2 X \Omega_{\up\up}\rangle}
\label{compare}
\eea
where $\tau_n$ denotes the translation over $n$ lattice units
in the chain. It follows that $\gamma_{\alpha\beta}\leq
\gamma_{\up\up}$.

For the proof of the upper bound it is convenient to present the
dense subspace of $\ran H_{\alpha\beta}$  formed by the vectors
of the form \eq{psi} in a slightly different way. Observe that
the spaces $\ker H^\mu_\Lambda\subset\Hs_{\alpha\beta}$ are decreasing
in $\Lambda$. Therefore, in order to assure that a certain $\psi$
belongs to $\ran H_{\alpha\beta}$, it suffices to check that
$\psi\perp\ker H^\mu_\Lambda$ for some suitable $\Lambda$.

We fix an interval $[1,n]$ and introduce the usual spin wave operators
$X_k$, $k=2\pi m/n, m=0,\ldots, n-1$, given by
\be
X_k=\frac{1}{\sqrt{n}}\sum_{x=1}^n e^{ikx} S^-_x
\label{Xk}
\ee
The normalization and the allowed values for $k$ are chosen such that
\be
\langle\Omega_{\up\up}\mid X_l^* X_k\Omega_{\up\up}\rangle=\delta_{k,l}
\label{ortho}
\ee
The vectors $\psi$ we need for the upper bound are linear combinations
of two spin waves, i.e. $\psi=(c_1 X_{k_1}+c_2
X_{k_2})\Omega_{\up\up}$. Due to \eq{ortho} we have $\Vert
\psi\Vert^2=\vert c_1\vert^2 + \vert c_2\vert^2$. For any pair of
distinct $k_1,k_2$, the coefficients $c_1, c_2$ can be chosen such
that $G_{[1,n]}\psi=0$, i.e. $\psi \perp \ker H^\mu_{[1,n]}$. This
follows from the fact that $\ker H^\mu_{[1,n]}$ contains exactly one
vector for each eigenvalue of $S^3_{[1,n]}$. All vectors
$X_k\Omega_{\up\up}$ have $S^3_{[1,n]}=(n-2)/2$. It follows that any
two-dimensional space of vectors $\psi$ with fixed, distinct $k_1,
k_2$ and arbitrary $c_1,c_2$ must contain a ray $\perp \ker
H^\mu_{[1,n]}$. Hence the upper bound \eq{Gammaupper} of Theorem
\ref{thm:main} can be proved by showing that
\be
\inf_{\mathstrut n,k_1,k_2}\sup_{\mathstrut c_1,c_2}
\frac{\langle \psi \mid H_{\up\up}\psi\rangle}{\langle
\psi\mid\psi\rangle}
=\Gammaupper(\Delta)\equiv 1- \frac{1}{\Delta}
\label{upperbound}
\ee
which we do next.

{}From the definition \eq{Xk} of the $X_k$ it is clear that the only
matrix elements of $H_{\up\up}$ we need are the $T_{x,y}$,
$1\leq x,y\leq n$, defined by
\be
T_{x,y}=\langle\Omega_{\up\up}\mid S^+_x H^\mu_{[0,n+1]}
S^-_y\Omega_{\up\up}\rangle
=\frac{1}{2\Delta}\left\{2\Delta\delta_{x,y}-\delta_{x,y-1}
-\delta_{x,y+1}\right\}
\label{Txy}
\ee
It is then easily seen that the $\sup_{c_1,c_2}$ in the left side of
\eq{upperbound} yields the norm of the $2\times 2$ matrix
$M(n,k_1,k_2)$ with matrix elements \be
M(n,k_1,k_2)_{i,j}=M_n(k_i,k_j)
\label{M}
\ee
where $M_n(k,l)$, for $k,l$ of the form $2\pi m/n$, is the function
\bea
M_n(k,l)&=&\frac{1}{n}\sum_{x,y=1}^n e^{-ikx}T_{x,y}e^{iky}\\
        &=&\delta_{k,l}(1- \Delta^{-1}\cos k)
            + (e^{il}+e^{-ik})/(2\Delta n)
\label{mkl}
\eea
It is then obvious that $\inf_{n,k_1,k_2}\Vert M(n,k_1,k_2)\Vert
= 1 -\Delta^{-1}$.

A similar calculation yields the $L$-dependent upper bound
\eq{isotropicgap} in the case $\Delta =1$.
\endproof

\section*{Acknowledgements}

It is a pleasure to thank the following people for discussions,
correspondence, and references: G.~Albertini, J.~Gr\"{u}neberg,
A.~Kl\"{u}mper, V.~Korepin, E.H.~Lieb, K.~Schoutens, J.-Ph.~Solovej,
and R.F.~Werner. We are  grateful to R.F.~Werner and W.F.~Wreszinski
for making their work (\cite{GW} and \cite{ASW}) available to us prior
to publication. The author is partially supported by the U.S. National
Science Foundation under Grant No. PHY90-19433 A04.

\def\thebibliography#1{\section*{References}\list
  {\arabic{enumi}.}{\settowidth\labelwidth{[#1]}
    \leftmargin\labelwidth
    \advance\leftmargin\labelsep
    \usecounter{enumi}}
    \def\newblock{\hskip .11em plus .33em minus -.07em}
    \sloppy
    \sfcode`\.=1000\relax}

\end{document}
\bye